\pdfoutput=1
\documentclass[cits]{PoS}
\usepackage{amsmath}

\def\sumint{\hbox{$\sum$}\!\!\!\!\!\!\!\!\int}

\title{Thermodynamics of the pion gas using the $\mathrm O(N)$ model in $1/N$ expansion}

\ShortTitle{$\mathrm O(N)$ model in $1/N$ expansion}

\author{\speaker{Tom\'{a}\v{s} Brauner}%
        \thanks{On leave from Department of Theoretical Physics, Nuclear Physics Institute ASCR,
                25068 \v{R}e\v{z}, Czech Republic.}\\
        Goethe-Universit\"{a}t Frankfurt am Main\\
        E-mail: \email{brauner@ujf.cas.cz}}

\abstract{We investigate the thermodynamics of a pion gas within the $\mathrm
O(N)$ model in the $1/N$ expansion. Using the auxiliary field technique, we
compute the effective potential up to the next-to-leading order (NLO) and show
that it can be renormalized in a temperature-independent manner. The crucial
step for the consistency of the calculation turns out to be the elimination of
the auxiliary field prior to renormalization. Subsequently, we solve the
NLO gap equation for the chiral condensate as a function of temperature both in
the chiral limit and with explicit symmetry breaking. We propose a simple
semi-analytic estimate of the NLO correction to the condensate and compare it
to the exact numerical solution. Finally, we show that in the chiral limit the
chiral symmetry is restored at finite temperature by a second-order phase
transition, and determine the critical scaling of the order parameter. We study
the dependence of the critical temperature on the renormalized coupling and
find that in contrast to the weak-coupling limit, at strong coupling the
critical temperature increases at NLO.}

\FullConference{8th Conference Quark Confinement and the Hadron Spectrum\\
                September 1-6, 2008\\
                Mainz. Germany}

\begin{document}

\section{Introduction}
The low-energy physics of Quantum Chromodynamics (QCD) is governed by
chiral symmetry, associated with the presence of (almost) massless quarks. In
the sector of the lightest $u$ and $d$ quarks, the spontaneous breaking of the
$\mathrm{SU(2)_L\times SU(2)_R}$ chiral symmetry in the vacuum gives rise to an
isospin triplet of pseudo-Nambu--Goldstone bosons, the pions. As the lowest
excitations of the ordered ground state, these dominate the thermodynamics at
low temperatures. Since the conventional perturbation theory breaks down at
finite temperature, one needs a suitable nonperturbative resummation scheme to
deal with the thermodynamics, in particular with the symmetry-restoring phase
transition. In QCD with two quark flavors, one can make use of the isomorphism
of the chiral group with the $\mathrm O(4)$ rotation group. The pions as well
as the sigma meson are described using the scalar $\mathrm O(4)$ model, which is
subsequently generalized to $\mathrm O(N)$, allowing for a nonperturbative
(one-particle-irreducible) $1/N$ expansion \cite{Coleman:1974jh}.

In this contribution we further develop the calculation of
\cite{Andersen:2004ae,Warringa:2006rn} at the next-to-leading order (NLO) in
the $1/N$ expansion. In particular we show that the NLO effective potential can
be renormalized by temperature-inde\-pen\-dent counterterms. This allows us to
renormalize and solve the NLO gap equation. From the methodical point of view,
our main conclusion is that for a consistent renormalization one has to dispense
with the auxiliary field introduced in \cite{Coleman:1974jh} (see
\cite{Jakovac:2008zq} for an alternative approach). From the physical point of
view, we show that the NLO truncation predicts a second-order chirally-restoring
phase transition and the critical behavior in accordance with general
universality arguments. For details of the calculations as well as a more
complete bibliographical account of related work we refer the reader to
\cite{Andersen:2008qk}.

\section{The auxiliary field technique}
At zero temperature and density, the $\mathrm O(N)$ sigma model is defined by
the Euclidean Lagrangian
\begin{equation*}
\mathcal L
=\frac12(\partial_\mu\phi_i)^2+\frac{\lambda}{8N}(\phi_i\phi_i-Nf_{\pi}^2)^2
-\sqrt{N}H\phi_N,
\end{equation*}
where $i=1,\dotsc,N$, and the parameters $f_{\pi}$ and $\lambda$ denote the
pion decay constant and coupling, respectively. The $H$-term accounts for
explicit symmetry breaking by nonzero quark mass. The essence of the
auxiliary-field trick \cite{Coleman:1974jh} is to add a new field $\alpha$ by
means of a pure Gaussian action,
\begin{equation*}
\Delta\mathcal L_\alpha=\frac
N{2\lambda}\left[\alpha-\frac{i\lambda}{2N}(\phi_i\phi_i -Nf^2_{\pi})\right]^2.
\end{equation*}
In case we wish to study pion gas at finite isospin density and pion
condensation \cite{Andersen:2006ys,Andersen:2007qv,Abuki:2008wm},
we further have to introduce the isospin chemical potential
$\mu_I$ in terms of the covariant derivative of the scalar field. Choosing
$\phi_{1,2}$ to represent the charged pion $\pi^\pm$, this amounts to adding
\begin{equation*}
\Delta\mathcal L_\mu=-i\mu_I
(\phi_1\partial_0\phi_2-\phi_2\partial_0\phi_1)-
\frac 12\mu_I^2(\phi_1^2+\phi_2^2).
\end{equation*}
As a next step one defines the chiral condensate,
$\phi_0\equiv\langle\phi_N\rangle/\sqrt N$, the pion condensate,
$\rho_0\equiv\langle\phi_1\rangle/\sqrt N$, and the condensate of the auxiliary
field, $iM^2\equiv\langle\alpha\rangle$. The fluctuations of the fields are
suppressed with respect to the condensates by the factor $1/\sqrt N$. This
ensures that the NLO contribution to the effective action is given by a simple
Gaussian integral over the fluctuations. The effective potential up to NLO thus
becomes
\begin{multline}
V_{\text{eff}}=\frac12NM^2(f_{\pi}^2-\phi_0^2-\rho_0^2)
+\frac{NM^4}{2\lambda}+NH\phi_0+\frac12N\mu_I^2\rho_0^2-\frac12N\sumint_P\ln(P^2+M^2)-\\
-\frac12\sumint_P\ln[I(P,M)]-\frac12\sumint_P\ln\left[(P^2+m^2)^2+4\mu_I^2\omega_n^2\right]
+\sumint_P\ln(P^2+M^2),
\label{effective_potential}
\end{multline}
where $m^2=M^2-\mu_I^2$, $\omega_n$ is the bosonic Matsubara frequency, and we
used a compact notation for the sum-integral over the four-momentum
$P\equiv(i\omega_n,\vec p)$. Also, we denoted
\begin{equation*}
I(P,M)=\frac1\lambda+\frac{\phi_0^2}{P^2+M^2}
+\frac{\rho_0^2(P^2+m^2)}{(P^2+m^2)^2+4\mu_I^2\omega_n^2}+
\frac12\sumint_Q\frac1{Q^2+M^2}\frac1{(P+Q)^2+M^2}.
\end{equation*}
The leading-order (LO) effective potential is given by the first line of Eq.
\eqref{effective_potential}. The sum-integral involved is divergent, but the
divergence can, after regularization using a four-dimen\-sio\-nal Euclidean
cutoff, be absorbed into renormalization of the parameters $f_\pi^2$ and
$1/\lambda$. The requirement of the stationarity of the effective potential
leads to a set of gap equations, whose solution yields the standard phase
diagram \cite{Andersen:2006ys}.

\section{Next-to-leading order}
The NLO effective potential is given by the second line of Eq.
\eqref{effective_potential}. The last two terms represent the correction to the
pressure of the free pion gas induced by a finite chemical potential. With a
four-dimensional cutoff (which is crucial for the successful renormalization at
the next-to-leading order), they give rise to an unphysical $\mu_I$-dependent
quadratic divergence. In the following, we therefore restrict ourselves to
finite temperature, but zero chemical potential.

\begin{figure}
\begin{center}
\includegraphics[scale=0.95]{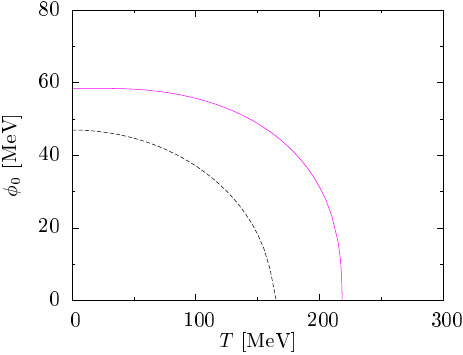}\hfill
\includegraphics[scale=0.95]{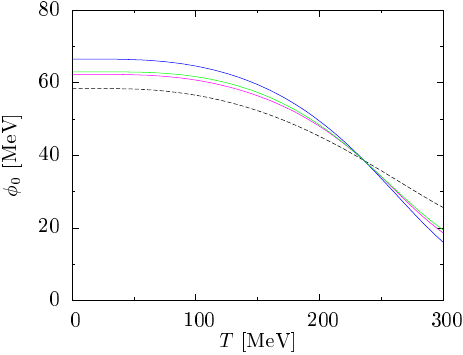}
\end{center}
\caption{Chiral condensate as a function of temperature in the chiral limit
(left panel) and at the physical point (right panel). LO (dashed line) and NLO
(solid red line) values are compared. For explanation of the green and blue
lines see the text.}
\label{Fig:NLO_condensates}
\end{figure}

In general, the NLO effective potential has to be evaluated numerically.
However, its divergent part can be calculated analytically
\cite{Andersen:2004ae,Warringa:2006rn}. The coefficient of the quadratic
divergence depends explicitly on temperature; this dependence disappears only
when the LO gap equation for $M$ is used. This suggests the necessity for a
reinterpretation of the effective potential \cite{Andersen:2008qk}. Since the
auxiliary field $\alpha$ was introduced as a composite operator of $\phi_i$, it
may give rise to unphysical features of the effective potential. Once it is
eliminated using its LO equation of motion, the resulting effective potential
as a function of $\phi_i$ alone can be renormalized in the usual manner.
Using this elimination procedure, we renormalized the NLO effective potential
and solved the gap equation for the chiral condensate. The results are shown in
Fig. \ref{Fig:NLO_condensates}. The numerical values were obtained with
$f_\pi=47\,\text{MeV}$ (differing by a factor $1/2$ from the conventional
value), $\lambda=30$, renormalized at the scale $100\,\text{MeV}$, and
$H=(104\,\text{MeV})^3$, adjusted in order to reproduce the physical pion mass
in the vacuum.

Apart from a direct extremization of the effective potential, one may attempt
to solve the gap equation by a systematic expansion in powers of $1/N$. Write
generally the effective potential per degree of freedom as $V(\phi)=V_0(\phi)+x
V_1(\phi)$, where $V_{0,1}$ denote the LO and NLO contributions. Apparently,
setting $x=0$ restricts the effective potential to its LO value, while
$x_0=1/4$ recovers the NLO result. Treating $x$ as a continuous variable, i.e.,
demanding that the gap equation $dV/d\phi=0$ be satisfied for all
$x\in[0,x_0]$, one derives a formal integral equation for the condensate,
\begin{equation*}
\phi(x_0)=\phi(0)-\int_0^{x_0}dx\,\frac{V_1'(\phi(x))}{V_0''(\phi(x))+xV_1''(\phi(x))}.
\end{equation*}
Assuming that the second derivative of the effective potential is constant
(that is, approximating it at the LO point by a parabola), we find the explicit
expression
\begin{equation*}
\phi(x_0)=\phi(0)-\frac{x_0V_1'(\phi(0))}{V_0''(\phi(0))+x_0V_1''(\phi(0))}.
\end{equation*}
This is the green line in Fig. \ref{Fig:NLO_condensates}. Note that the formal
$1/N$ expansion to the next-to-leading order would require neglecting $V_1''$ in
the denominator (blue line in Fig. \ref{Fig:NLO_condensates}). Obviously this
is not a good approximation: It overshoots the NLO correction to the chiral
condensate by nearly $100\%$.

\begin{figure}
\begin{center}
\includegraphics[scale=1]{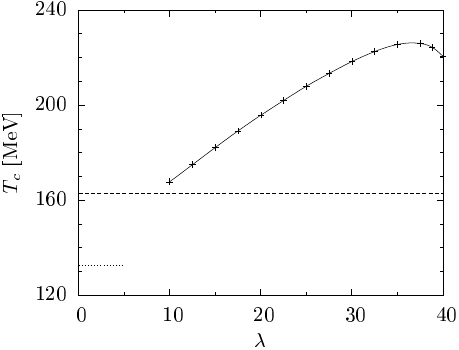}
\end{center}
\caption{NLO critical temperature as a function of the coupling. The dashed
line denotes the LO (coupling-independent) value, while the dotted line
indicates the NLO weak-coupling limit.}
\label{Fig:Tc}
\end{figure}
In Fig. \ref{Fig:Tc} we display the dependence of the NLO critical temperature
on the renormalized coupling. While the weak-coupling analytic calculation
predicts a negative correction, $T_c=\sqrt{\frac{12}{1+\frac2N}}f_\pi$, the
numerical results show that at strong coupling, the critical temperature is
increased by the NLO correction. Even though the auxiliary-field technique does
not allow us to check the weak-coupling limit directly, the decreasing trend
towards smaller values of the coupling is reassuring. Finally, from the solution
of the gap equation near the critical temperature we were able to extract the
critical exponent, which governs the scaling of the order parameter. The
numerical value is in agreement with the analytic NLO expression,
$\nu=\frac12-\frac4{N\pi^2}$.

To conclude, we have shown that the NLO effective potential of the $\mathrm
O(N)$ model can be consistently renormalized in a temperature-independent
manner provided one uses the LO equation of motion to eliminate the auxiliary
field before renormalization. We used this strategy to solve the NLO gap
equation and proposed a simple semi-analytic approximation to the NLO chiral
condensate which is very close to the exact value. We also investigated the
coupling-dependence of the NLO critical temperature and found that the NLO
correction is negative at weak coupling, but positive at strong coupling. In
general, the NLO corrections to the observables turn out to be consistent with
the $1/N$ expansion, i.e., roughly of the order of ten percent.

\acknowledgments The author is grateful to J. O. Andersen for collaboration
which gave rise to the results presented here. This work was supported in part
by the Alexander von Humboldt Foundation and by the GA CR Grant No. 202/06/0734.

\providecommand{\href}[2]{#2}\begingroup\raggedright\endgroup

\end{document}